# Edge Computing in Transportation: Security Issues and Challenges

Nikheel Soni[1,2], Reza Malekian[2,3], Arnav Thakur[2,4]

[1]Amazon Web Services, Cape Town, 8001, South Africa
[2]Department of Electrical, Electronic and Computer Engineering, University of Pretoria, Pretoria, 0002, South Africa
[3] Department of Computer Science and Media Technology, Malmö University, Malmö, 20506, Sweden
[4] Developer and Mobile Services, Amazon Web Services, Cape Town, 8001, South Africa
Reza.malekian@ieee.org

*Abstract* — As the amount of data that needs to be processed in real-time due to recent application developments increase, the need for a new computing paradigm is required. Edge computing resolves this issue by offloading computing resources required by intelligent transportation systems such as the Internet of Vehicles from the cloud closer to the end devices to improve performance however, it is susceptible to security issues that make the transportation systems vulnerable to attackers. In addition to this, there are security issues in transportation technologies that impact the edge computing paradigm as well. This paper presents some of the main security issues and challenges that are present in edge computing, which are Distributed Denial of Service attacks, side channel attacks, malware injection attacks and authentication and authorization attacks, how these impact intelligent transportation systems and research being done to help realize and mitigate these issues.

## I. INTRODUCTION

With the ever-increasing amount of data that is being generated and processed today, it is extremely critical that the upmost security best practices and standards are maintained in order to protect this data. Approximately 2.5 quintillion bytes of data is generated per day by the internet [1]. This is mainly due to a rapid increase in the number of devices that are being connected to the internet, more well known as the Internet of Things (IoT). IoT has allowed us to manipulate and control devices we use every day such as adjusting the ambient lighting in smart buildings to sharing data between vehicles to improve applications such as autonomous driving. Looking towards improving vehicle technology development, the focus of this paper is on the Internet of Vehicles.

With the recent advances in vehicle technology, vehicles are no longer just simply consist of an engine with an accelerator and brake pedal. Internet of Vehicles, otherwise known as IoV, has allowed for vehicles to gather data from various sensors in a vehicle which include, but are not limited to, GPS data, vehicle speed, applied braking power and even tire pressure to create applications such as collision avoidance systems, traffic routing and smart parking systems as well as improve air quality through more efficient driving [2] to be used in a smart vehicle network. These applications require an extensive amount of data to be generated by the sensors on board a vehicle which need to be further analyzed.

An example of how much data that can be produced from a vehicle can be explicitly seen from Formula 1 (F1). From various sensors on an F1 car, approximately 3 GB of data, and 1,500 data points are generated each second [3]. This data is then transmitted to both the pit wall and a team's home factory which can be in another country from where a race is being held [4] in order to analyse and make decisions that are crucial to winning a race. While the technology that F1 cars use is not used on public roads, the extensive amount of research and development that is being done is allowing for some of these technologies to be introduced into road legal vehicles which will help improve performance and fuel emissions of a vehicle [5].

Due to the amount of data that can be generated by sensors on a vehicle, the onboard systems tend to lack the capacity to do this due to computational limits. Thus, a critical component of the IoV paradigm is in fact cloud computing. While cloud computing has allowed us to provision resources that are essential for IoV and other IoT solutions to operate on, due to its relatively centralized design, it is susceptible to flaws such as single point of failure, increased latency due to data propagation, network connectivity dependency due to geographical location accessibility and security and privacy concerns to mention a few [6]. Because of this, it is essential that we attempt to offload and distribute the computational power and resources required by IoV infrastructure to the edge of the network closer to end devices and users. This can be accomplished by introducing a resource layer between cloud computing and IoT devices called edge computing.

It is technology such as edge computing that can help drive the innovation of that created in the IoV paradigm by allowing data to be processed closer to the source for improved performance in data processing and analysis required by applications such as dynamic traffic routing for vehicles. Bringing intelligence to traffic analysis and management can help decrease the amount of time a driver spends in traffic, decrease fuel emissions, and reduce traffic density [7]. In addition, edge computing will allow for other various intelligent transportation solutions such as vehicle tracking systems and public transportation management to operate in a near real-time environment which will improve the overall operational efficiency.

Introducing an edge computing layer into IoV and cloud computing systems will prove beneficial to performance, real-time data delivery as well as improved data privacy however, it also introduces additional security risks and challenges to the infrastructure on top of those that already exist in IoV and cloud based systems. This is because the edge computing infrastructure makes itself vulnerable due to reduced compute power, heterogenic devices and operating systems and attack unawareness [8]. The main aim of this paper is to discusses the most influential security issues and challenges that currently exist within edge computing in transportation and how these issues can impact IoV systems. Furthermore, we briefly discuss security issues and challenges in IoV systems and how they can essentially magnify the issues in edge computing.

The remaining of this paper is structured as follows. In section II we present the architectures of both edge computing and IoV and how they are interrelated with each other. In section III, we discuss research that has been done with respect to vehicle attacks. In section IV we discuss the security issues present in edge computing and what research has been done on an IoV level for these issues. In section V, we briefly discuss the reason for the security issues mentioned in section IV and finally, the paper is then concluded in section VI.

## II. ARCHITECTURE

To understand where the security issues and vulnerabilities lie in both edge computing and IoV systems, we need to look at how each paradigm is architected. Firstly, we look at edge computing and thereafter, we look at IoV and how it integrates into edge computing.

### A. Edge Computing Architecture

The main goal of an edge computing architecture is to bring computational power and resources closer to the end user. For this, a typical edge computing architecture is composed of 3 different layers as depicted in figure 1.

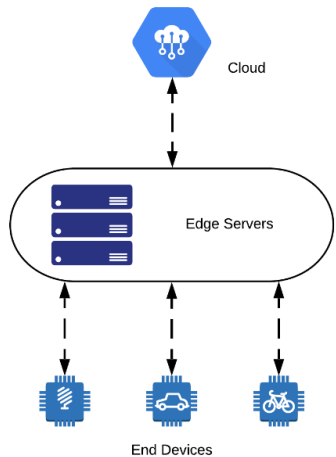

Fig. 1. Basic Edge Computing Architecure

Before the addition of the edge layer, end devices were connected directly to the cloud. While this 2-tier based model may have worked efficiently during its inception, with the ever number of increasing devices being connected to the cloud, an increase in latency, computational load and last mile communication to a cloud server to name a few, has increased. Furthermore, adding security measures to each of these communication streams adds an excessive amount of network overhead [9]. To help reduce the strain on the cloud, the introduction of an additional layer called the "Edge" has been presented which can be seen in figure 1. The Edge layer essentially resides between end devices and the cloud which consists of edge computing servers. Because edge computing is distributed by design, its availability for resources required by that of end devices is improved in cases where the cloud is not available to an end user.

The main goal of an edge computing server is to reduce the load on a cloud server, improve latency and provide a high QoS [10] in an addition to improving privacy and security as data is closer to the end user as it does not need to traverse such a great distance. Edge servers will serve as the first contact point for end devices when computational power and resources are required. If an edge device is not capable of serving a given request from an end device, it is forwarded to the cloud [10]. A typical example of where edge computing is essential is in the vehicle technology sector where real-time data analysis is crucial to vehicle applications such as traffic analysis [11] and autonomous driving [12].

### B. IoV Architecture

With the foundation set for Edge Computing architecture, it is relatively easy to see how IoV infrastructure integrates into edge computing. A typical example of an IoV architecture [13] can be seen in figure 2.

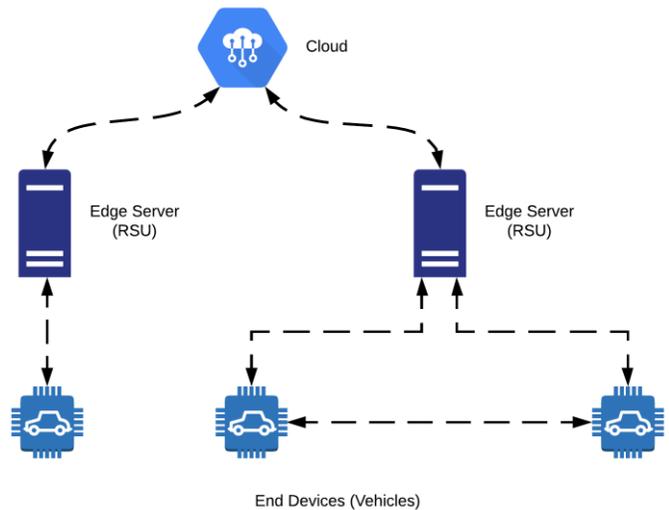

Fig. 2. IoV Architecture

In this architecture, we have the vehicles acting as the end devices that are generating data from various on-board sensors on the vehicle used to determine external conditions. The data being generated is processed locally on board the vehicle when possible and thereafter, is passed to either an edge server for further analysis or directly to another vehicle. Aligning this architecture with edge computing, there are two main modes for communication in IoV systems namely vehicle-to-vehicle (V2V) and vehicle-to-infrastructure (V2I). With respect to V2I, this form of communication is done between a vehicle and what is commonly referred to as a roadside unit (RSU), which can be considered as an edge computing server. Because of the constant geographical changes in a vehicles position, a vehicle may connect with several different edge servers along a single trip in order to obtain information about the new events in the region of the newly connected edge server with the same being said for V2V communication making a IoV network highly dynamic in nature.

As it can be seen, the vulnerabilities that exist in edge computing are essentially inherited by that of the IoV paradigm and those issues that are in IoV can be exploited and impact edge and cloud environments.

## III. STATE OF THE ART

The need for improved security measures in edge computing and IoV can be seen by viewing recent research being done with respect to vehicle attacks. This has first started off from attacks happening when physical access to a vehicle is possible to more advanced attacks where an attacker does not need to be within any defined range and can be hacked from any location where a cellular connection is present.

In [14], tests were performed in a lab and closed road environment which showed that if an attacker has physical access to a vehicle, the damage could be catastrophic to both the driver as well as the technology used within the vehicle. With physical access to the vehicle, [14] shows the vulnerabilities within a car that can be exploited. Because of the lack of authentication to the CAM bus, once an attacker is physically connected to the vehicle, an attacker was able to manipulate various parts of the vehicle that were critical to controlling it such as the data shown on the instrument cluster, change engine speed and enabling individual brakes on the vehicle. Other smaller attacks included interfering with the infotainment system, adjusting the lighting throughout the vehicle and toggling door locking mechanisms which can prove as distractions to a driver and divert their attention away from the road.

Furthermore, traces of these attacks can be wiped from the vehicles logging system meaning that it will be impossible to investigate where the attack came from and what was impacted.

While [14] expresses that the hacker be on site to perform any malicious activity to a vehicle, the work in [15] took this one step further by remotely hacking a vehicle. Researchers managed to hack a Jeep Cherokee using either a Wi-Fi or cellular connection to gain access to the vehicles infotainment system. Once this was accomplished, access to send commands to the CAM bus was obtained by performing a firmware upgrade. This allowed for the hackers to control various functions of the vehicle, which included the steering, brakes and engine transmission to mention a few. This vehicle hack has shown that there are not only wireless access vulnerabilities and security issues present, but also software design flaws between entities such as the infotainment system and CAM bus, which should be essentially sandboxed to prevent malicious access.

In [16], researchers managed to find and exploit 14 security vulnerabilities found in several BMW vehicles. These attacks ranged from having direct access to the vehicle to remotely hacking the vehicle. This prompted the German manufacturer to provide security patches over-the-air and via their dealer networks. While the vehicles in question are not autonomous, having access to the infotainment system and CAN BUS can exploit the vehicles operational expectations.

Authors in [17] demonstrated how we do not need to have physical or remote access to a vehicle at all however, if the technology that is used in the vehicle has vulnerabilities, it is possible to exploit them. Researchers from the McAfee advance research team managed to deceive the Mobileye EyeQ 3 used in Tesla vehicles to make it think that a 35 mile per hour road sign was actually a 85 mile per hour road sign by manually changing the road sign with a piece of electrical tape. This type of attack shows that machine vision algorithms can sometimes provide incorrect information obtained from its surroundings.

In addition to the increasing amount of research that is being conducted proving that security flaws exists in automobiles, various cloud and technology companies such as Amazon and BlackBerry [18], Daimler and Bosch [19] and Microsoft and Toyota [20] among others are starting to collaborate together on the path to delivering autonomous vehicle technology. Furthermore, companies such as Cisco plan to bring gigabit-speed connectivity to smart vehicles [21], which can widen the scope of an attack on a smart vehicle due to increased bandwidth capabilities being delivered to a vehicle. Higher bandwidth capabilities are crucial to autonomous vehicles as a report from Intel states that 1 GB of data will need to be processed per second in real-time for a vehicle to make a decision based on its environment captured from sensor data [22].

## IV. SECURITY RISKS & CHALLENGES

As it can be seen in the architecture for both edge and IoV paradigms, security threats and risks that are present in edge computing are inherited by IoV systems and vice versa. In this section, we will discuss four of the most impacted security vulnerabilities in edge computing and the research being done in each of these areas with regards to IoV. As there are several different security flaws and issues present in edge computing systems, we will only focus on the four most impacted vulnerabilities as discussed by [7], which provides an extensive overview of these security vulnerabilities. This includes Distributed Denial of Service attacks (DDoS), side-channel attacks, malware injection attacks and authentication and authorization attacks.

### A. Distributed Denial of Service Attacks

In a DDoS attack, an attacker targets a server to bring it to a halt which essentially, denies service availability to those that require the service that the server is providing. This is done by flooding a server with an abundant number of requests so that it exhausts all possible resources on a server such as compute, memory and networking. While there are different types of DDoS attacks, these can be categorized in application level attack, protocol level attack and volume-based attacks [23].

In [24, 25], an approach to detect and prevent DDoS attacks is done by analysing the communication time to a node in a vehicular network. If the communication time is beyond a certain threshold that does not conform to the set communication rules, it will stop communicating with the attacking source as well as inform all other nodes the details of the attacker. A similar approach is done in [26], where the network bandwidth of nodes is monitored to see if set thresholds are breached and if so, remove the compromised node from the network.

*B. Side Channel Attacks*

In a side-channel attack, an attacker makes use signals generated such as those of a power supply or communication [27] and analyses them in order to gain information required to access a system. Because the information that the attacker is using such as power supply signals which is open publicly, it is quite difficult to prevent an attacker from obtaining this information.

Side-channel attacks have not been researched in-depth with regards to IoV systems and as a result, prove to be more challenging to defend against. In [28], researchers have shown that analysis of side channel information using fuzzy logic can infer the secret bits of a cryptosystem which can be used to intercept a data connection on various vehicle embedded systems such as the ECU or infotainment system.

*C. Malware Injection Attacks*

A malware injection attack is when an attacker injects malicious software on a device that compromises is system integrity. Once an attacker has injected their malware in an edge server, various malicious acts can be performed such as damaging a back end database to consuming system resources which impacts system performance and even encrypt data and hold it as a hostage in the form of ransomware.

While, various ways to defend against malware have been investigated such as signature and behaviour-based detection, these detection methods may not be feasible to implement in vehicle systems due to database storage limitations and a lack of compute power [29]. The works presented in [30] and [31] introduce frameworks to help improve defences in malware attacks. In [30], an external entity is added to the communications of the vehicle which is referred to as a security cloud which analyses data that the onboard system cannot verify if malicious or not in addition to analysing incoming network traffic to the vehicle while in [31, 32], a virtualization approach is followed which reduces the number of computational components required in a vehicle thus, decreasing the attack surface for malicious activity where each component will have processes in place for malware prevention and detection.

*D. Authentication and Authorization Attacks*

When looking at access and permissions on for servers and services, our attention focuses to authentication and authorization. Authentication refers to who has access to a particular entity whereas authorization refers to what actions an authenticated entity can perform when access is granted. Attacks that are authentication-based include dictionary attacks and attacks exploiting vulnerabilities in authentication protocols whereas attacks that are authorization-based include attacks exploiting vulnerabilities in authorization protocols and overprivileged attacks [23].

In [33], research was done on using aggregating message authentication codes to ensure that a message transmitted within an IoV network can be verify for authenticity and integrity while keeping communication overheads to a minimum. In [34], various security requirements such as authentication, availability, message confidentiality, message integrity, data availability, access control, privacy, message non-repudiation and real-time guarantees of message delivery is discussed which is highly important in improving authentication and authorization mechanisms.

In all of these attacks, edge computing can be highly susceptible due to the fact that end devices and servers are relatively less powerful than cloud based systems and are heterogenic in nature [7] therefore, do not have high security measures in place to defend against these attacks. Furthermore, if a single node is compromised from a device layer, being a vehicle in this situation, the attack could first spread laterally to other devices. These devices can then be used to bring down edge servers, which could in turn bring down the cloud environment through a DDoS attack that has been spawned on various edge servers and devices.

V. DISCUSSION

As it can be seen security is a major concern when it comes to edge computing and IoV paradigms. It is thoroughly discussed in [7] that the main issues in edge systems are due to either protocol or implementation flaws with the exception to side-channel attacks. Furthermore, when systems are designed, little effort is focused on areas such as security, whereas more effort is focused on user interaction and experiences. Bringing devices into the infrastructure only increases the attack surface that an attacker can breach the edge and IoV infrastructures from. When looking at IoV security, one can think of the basic security requirements that should be incorporated into the design of these systems, which include authentication, data integrity, non-repudiation, real-time availability and confidentiality. Heterogeneity is also of a major concern with regards hardware, protocols and software implementations used within the edge and IoV paradigm as there is no common defense that can protect all these factors therefore, the need for a suitable framework needs to be developed and adopted by devices manufactures and edge and cloud providers to help mitigate this.

VI. CONCLUSION

This paper discussed the some of the main issues that are present in edge computing due to is design flaws and how theses issues can impact end devices. From the state of the art, we identified the need to improve security in transportation systems is critical to both end users as well as underlying system operability. Further research is required in both edge computing and IoV paradigms with regards to improving security flaws that exist without compromising application performance and user interaction. This can be done by implementing a suitable framework to conform to set standards and eliminate heterogeneity within these systems.